# Constraints on the Production of Phosphine by Venusian Volcanoes

**William Bains** [1,2,*], **Oliver Shorttle** [3,4], **Sukrit Ranjan** [5,6,7], **Paul B. Rimmer** [3,8], **Janusz J. Petkowski** [1], **Jane S. Greaves** [2] **and Sara Seager** [1,9,10]

[1] Earth, Atmospheric and Planetary Sciences, Massachusetts Institute of Technology, Cambridge, MA 02139, USA; jjpetkow@mit.edu (J.J.P.); seager@mit.edu (S.S.)
[2] School of Physics & Astronomy, Cardiff University, 4 The Parade, Cardiff CF24 3AA, UK; GreavesJ1@cardiff.ac.uk
[3] Department of Earth Sciences, University of Cambridge, Downing Street, Cambridge CB2 3EQ, UK; os258@cam.ac.uk (O.S.); pbr27@cam.ac.uk (P.B.R.)
[4] Institute of Astronomy, University of Cambridge, Madingley Rd, Cambridge CB3 0HA, UK
[5] Department of Astronomy & Astrophysics, Northwestern University, Evanston, IL 60201, USA; sukrit@mit.edu
[6] Center for Interdisciplinary Exploration and Research in Astrophysics, Northwestern University, Evanston, IL 60201, USA
[7] Blue Marble Space Institute of Science, Seattle, WA 98154, USA
[8] Cavendish Laboratory, University of Cambridge, JJ Thomson Ave, Cambridge CB3 0HE, UK
[9] Department of Physics, Massachusetts Institute of Technology, 77 Mass. Ave., Cambridge, MA 02139, USA
[10] Department of Aeronautics and Astronautics, Massachusetts Institute of Technology, 77 Mass. Ave., Cambridge, MA 02139, USA
* Correspondence: bains@mit.edu

**Abstract:** The initial reports of the presence of phosphine in the cloud decks of Venus have led to the suggestion that volcanism is the source of phosphine, through volcanic phosphides ejected into the clouds. Here, we examine the idea that mantle plume volcanism, bringing material from the deep mantle to the surface, could generate observed amounts of phosphine through the interaction of explosively erupted phosphide with sulfuric acid clouds. The direct eruption of deep mantle phosphide is unphysical, but a shallower material could contain traces of phosphide, and could be erupted to the surface. The explosive eruption that efficiently transports material to the clouds would require ocean:magma interactions or the subduction of a hydrated oceanic crust, neither of which occur on modern Venus. The transport of the erupted material to altitudes coinciding with the observations of phosphine is consequently very inefficient. Using the model proposed by Truong and Lunine as a base case, we estimate that an eruption volume of at least 21,600 $km^3$/year would be required to explain the presence of 1 ppb phosphine in the clouds. This is greater than any historical terrestrial eruption rate, and would have several detectable consequences for remote and in situ observations to confirm. More realistic lithospheric mineralogy, volcano mechanics or atmospheric photochemistry require even more volcanism.

## 1. Introduction

Greaves et al. [1] reported detecting ~20 ± 10 ppb phosphine ($PH_3$) in the atmosphere of Venus, using the James Clerk Maxwell Telescope (JCMT) and Atacama Large Millimeter/submillimeter Array (ALMA) radio telescopes. Subsequent to the publication of their paper, a recalibration of ALMA data and reprocessing the JCMT data [2] led to a revision of the abundance to 5–7 ppb with 5–6 $\alpha$ confidence. Infrared observation made by the SOIR (solar occultation in the infrared) instrument on Venus Express suggests a much lower upper limit abundance of <0.5 ppb above the clouds, at 60 km [3]. Other groups challenged the detection on technical grounds, suggesting that the detected line was artefactual [4,5] or was contamination by $SO_2$ [6], both of which were rebutted in [7,8]. A reanalysis of Pioneer Venus mass spectrometry data supports the presence of $PH_3$ in the clouds [9], with masses detected which are not consistent with any other species. Thus, two independent techniques that were applied in different decades suggest, but do not conclusively prove, the presence of PH3.

If PH3 is present in Venus' atmosphere, it is worth exploring why it might be there. A wide range of potential sources have been suggested, both in the literature [10] and in informal discussions. Bains et al. [11,12] discuss these at length, emphasizing that potential sources must be shown to quantitatively account for the presence of at least 1 ppb PH3 at cloud level in Venus' atmosphere to explain the detections. They show that, within the limits of the knowledge of the chemistry and photochemistry of the atmosphere of Venus, there is no known abiotic atmospheric source for PH3.

The chemistry and mineralogy of the surface and subsurface of the planet are less well known, but [11] show that thermodynamics make a surface volcanic source unlikely. They explored which gases could erupt on Venus with different temperatures, water abundances, and rock reduction regimes. Specifically, they show that, even if the surface was as reduced as the QIF redox buffer (implying the presence of metallic iron in erupted material), only ~1 part in $10^6$ of phosphorus in the erupted material would be present as phosphine. They also state that plume volcanism is unlikely, given a semi-quantitative argument. They comment that "on Earth, mantle plume magma is estimated to rise on a timescale of $10^6$–$10^7$ years at temperatures in excess of 3000 K, during which time phosphorus species would reach thermodynamic equilibrium relevant to the temperature and pressure of the upper mantle." This temperature and this pressure strongly disfavor phosphine.

Ref. [11] shows that a biological source in the clouds is not ruled out by thermodynamics, but comment that Venus' clouds are an extremely challenging environment due to high acidity and very low water activity, some of which have been partially addressed in follow-up work [13–15].

Subsequently, Truong and Lunine [16] revisited the concept that plume volcanism could be a source of PH3. Specifically, they propose a model in which plume volcanism could bring the P(-3) species (presumed to be metal phosphides) present in the deep mantle to the surface; explosive volcanism could lift them to the cloud layer, and the sulfuric acid in the clouds could then react with phosphides to make PH3.

In this paper, we examine the volcanic supply of phosphine scenario and determine under what conditions it could apply. We conclude that it would require a highly unexpected set of conditions for the model to be a valid source of PH3. The only possible scenario would require Venus to be entering a global resurfacing epoch, such as that proposed to have occurred ~300 Mya [17], with highly unexpected lithospheric chemistry, volcano dynamics, and atmospheric processes. Evidence of this could be found in a range of past and planned observations of Venus.

## 2. Materials and Methods

### 2.1. Phosphorus Partition between Iron and Silicate

The partition coefficient of phosphorus between iron and silicate in mixed iron/silicate melts ($D_P$) was calculated following [18] using Equation (1)

$$D_p = a \cdot \ln[f(O_2)] + \frac{b}{T} + \frac{c \cdot T}{P} + d \cdot \frac{nbo}{t} + f \qquad (1)$$

where $a$, $b$, $c$, $d$ and $f$ are constants, such that $a$ = 0.404, $b$ = 17455, $c$ = 579, $d$ = −0.725, $f$ = −3.15, $f(O_2)$ is the oxygen fugacity, $T$ is the absolute temperature, $P$ is pressure in gigapascals, and $nbo/t$ is the ratio of non-bridging oxygens to tetrahedral oxygens, which is related to the degree of polymerization of the silicate glass structure; in [18] $nbo/t$ ranged from 0.88 to 2.57, and was taken to be 1.5 here. Pressure as a function of depth was taken from the Preliminary Reference Earth model [19], corrected for Venus gravity (0.88 at the surface). The temperature of the Venusian mantle is poorly constrained. We have taken the average temperature profile for Earth from [20] as a plausible estimate. We note that the solubility of phosphorus in melts is an inverse function of temperature [18], and so if mantle plumes are hotter than the surrounding rocks, which is likely [21], this model will overestimate the partition of phosphorus into iron.

*2.2. Phosphorus Thermodynamics in Silicate Melts*

The thermodynamics of phosphorus speciation between iron phosphide and mineral phosphate in silicate melts were calculated as described in [11]. In brief, Gibbs free energy values were obtained from [22–26], and the density of minerals was taken from The Minerals Project [27]. The free energy of a reaction of reactants to give products was calculated using Equation (2)

$$\Delta G^r_{T,P} = \sum_{n=1,p} (S_{n.} \cdot \Delta G^T_n) - \sum_{n=1,r} (S_{n.} \cdot \Delta G^T_n) + \sum_{n=1,p} \left(\frac{S_{n.} \cdot V_n \cdot P}{10}\right) - \sum_{n=1,r} \left(\frac{S_{n.} \cdot V_n \cdot P}{10}\right) \quad (2)$$

where $\Delta G^r_{T,P}$ is the free energy of the reaction at temperature $T$ (in Kelvin) and pressure $P$ (in bar), $S_n$ is the stoichiometric number of moles of reagent $n$, $\Delta G^T_n$ is the free energy of formation of reagent $n$ from its elements at temperature $T$, $V_n$ is the molar volume of reagent $n$, $p$ are products, and $r$ are reactants. The physical state of the reactants is taken as a solid or a liquid depending on temperature, except for $O_2$, which is a gas throughout. $SiO_2$ was assumed to be quartz and metals were assumed to be in their lower oxidation state where relevant (i.e., Fe(II), not Fe(III) etc.). The free energy of reaction $\Delta G^r_{T,P}$ per mole of oxygen is converted to an oxygen fugacity (i.e., the activity of $O_2$ ($\{O_2\}$) at which the reaction is at equilibrium when the activity of oxidized and reduced iron species are the same) according to Equation (3)

$$f(O_2) = LOG_{10}(\{O_2\}) \sim \frac{-\Delta G^r_{T,P}}{R \cdot T \cdot 2.303} \quad (3)$$

The oxygen fugacity of QIF, IM and FMQ buffers, as a function of temperature and pressure was taken from [28]. The activity of free phosphorus in solution in iron at <10% phosphorus is negligible [24], and so, all the phosphorus can be considered to be present as iron phosphide, or as phosphate. Details of the reactions and of the calculation of the free energy of reaction and oxygen fugacity are provided in the Supplementary Materials.

*2.3. Phosphorus and Oxygen Diffusion through Solids*

The diffusivity of an element in a solid can be approximated by an Arrhenius Equation (4)

$$D_o = A \cdot e^{\frac{Ea}{R \cdot T}} \quad (4)$$

where $D_o$ is the diffusion constant (in this paper in units of $cm^2/day$), $R$ is the gas constant, $T$ is the absolute temperature, and constants $A$ and $Ea$ are as listed in Table 1.

**Table 1.** Constants for equation 4 to calculate diffusion coefficient.

| Element | Solid | A ($cm^2$/Day) | $Ea$ (kJ/mol) | Notes | Reference |
|---|---|---|---|---|---|
| P | Iron | $9.84 \cdot 10^4$ | 306 | Calculated from data in paper | [29] |
| O | Iron | $2.396 \cdot 10^{-1}$ | 118 | Minimally affected by dissolved Si, Al or Mg | [30] |

We note that the values for phosphorus diffusion through iron were only measured between 1050 °C and 1200 °C, and so, values outside this range represent an extrapolation. The rate of diffusion of phosphorus out of an iron sphere assumes a zero concentration outside the sphere, and the diffusion of oxygen into the sphere assumes a constant external concentration and no oxygen inside the sphere at the start. These were estimated by numerical integration; the result can be summarized by Equation (5)

$$t_{½} = \frac{r^2}{K \cdot D_o} \tag{5}$$

where *K* is a constant = 0.376 for diffusion of phosphorus out, and 0.355 for diffusion of oxygen into the sphere.

## 3. Results

We describe several constraints on the production of phosphine by Venusian volcanoes, and determine quantitatively under what conditions Venusian volcanism could be a valid source of $PH_3$. We build on the baseline case of [16] using the approach described in [11]. We first discuss why the mineralogy (chemical species) of erupted material is necessarily determined by the upper mantle and lithospheric mineralogy (Section 3.1), which leads to the detailed modelling of the abundance of phosphides in the mantle and lithosphere (Section 3.2). We then discuss why explosive volcanism is likely to be rare on Venus, (Section 3.3), why all volcanism is less likely to produce large ash clouds on Venus (Section 3.4), and some other limits on the production of phosphine from phosphide (Section 3.5), before reviewing the rate at which phosphide must be delivered to the clouds (Section 3.6).

Section 4 integrates these calculations to show that the minimum rate of eruption that explains 1 ppb phosphine in the clouds is 21,600 km³/year, not 0.03 km³/year as suggested by [16], even given the most optimistic assumptions.

*3.1. Eruptive Melt Mineralogy is Determined in the Upper Mantle and Lithosphere*

Before addressing the likely mineralogy of an erupting volcano, we briefly discuss why the mineralogy of that volcano is determined by the prevailing conditions of pressure, temperature and oxygen fugacity in the lithosphere and upper mantle, not the deep mantle.

We note that the minerology of a rock (the chemical species present, such as phosphate or phosphide) is distinct from the elemental abundance of a rock (the relative number of atoms of the elements in a rock). An assemblage of phosphorus, hydrogen and oxygen atoms could be present as phosphate and $H_2$, or as phosphide and $H_2O$. The elemental abundance depends solely on the source of the materials that made up the rock. The minerology—the chemical forms that those elements take—depends on the elemental abundance and on the pressure and temperature of the rock. In this paper, we are primarily concerned with the minerology of materials, not their elemental composition, which is assumed to be similar to Earth.

It is important to emphasize that oxygen fugacity is both directly a function of temperature, pressure, and rock composition, and indirectly affected by the effect of temperature and pressure on the mineralogy that is stable under a set of conditions (e.g., [31]). Oxygen fugacity is a property of the mineralogy of a rock, because it is the equilibrium position of a hypothetical reaction

$$MO \leftrightarrow M + O_2 \tag{6}$$

where MO is an oxidized mineral, M is a reduced mineral, and $O_2$ is molecular oxygen. The oxygen budget of the rock will of course be the same at low pressure as at high pressure, but the $f(O_2)$ that the abundance of O translates to will be dominated by the mineralogy attained by the sample at a specific temperature and pressure. The oxygen fugacity of a system therefore changes with changes in temperature and pressure [28], i.e., with depth in a planet, independently of any change in the composition of the system. Thus, the oxygen fugacity of a rock of fixed bulk composition at the base of the lithosphere will be substantially different from the oxygen fugacity of the same rock in the deep mantle, even though they have the same intrinsic oxidizing capacity and the same elemental composition. Therefore, for the oxygen fugacity of the deep mantle to dictate the composition of volcanism, which necessarily occurs at low pressure, the relaxation of a system to low pressure thermochemical equilibrium (and its associated higher $fO_2$) must be suppressed.

The deep mantle has been suggested, primarily on the basis of diamond inclusion chemistry, to be reducing [32]. Mantle convection, including mantle plumes, lifts this material to the base of the lithosphere, where it can pool and form melts that are subsequently erupted. A mantle plume is a buoyant upwelling of abnormally hot rock within the Earth's mantle [33]. The 'head' of a plume forms a mushroom-shaped bolus which establishes the channel behind it by pushing aside the cooler, denser, more viscous material of the mantle [34,35]. Such plumes have been suggested as dredging xenoliths from the 250 km-deep cratonic keels of continents [36].

While deep mantle mineralogy can be reflected in inclusions in highly refractile xenoliths, the xenoliths are not representative of the bulk erupted material from plume volcanism, which is overwhelmingly basaltic, for two key reasons. First, although mantle material may have upwelled from a great depth, its melting typically begins in the upper mantle, and certainly its volumetrically dominant melting occurs there. Therefore, at the point at which melting has occurred, the material will have an upper mantle mineral assemblage, which produces more intrinsically oxidizing conditions (for the same bulk oxygen content of the rock), as noted above. Second, even melts that are produced at great depth must transit through the overlying mantle and lithosphere. On Earth, melt transport out of the mantle dominantly occurs from the porous flow along grain boundaries, which brings the melt into close contact with progressively lower pressure mineral assemblages. The geometry of melt flow out of the mantle, and the high temperatures at which this is occurring, both promote the equilibration of melts surrounding rock. For this reason, even the largest outpourings of magma on Earth, which have been supplied by deep upwellings of mantle rock, evidence depths of last equilibration in the upper mantle.

The rate at which the geometry of melt transport brings the rock and fluid into equilibrium can be illustrated as follows. Iron is expected to be solid at any depth in the mantle [37], and as deep mantle phosphides are expected to be almost exclusively dissolved in the iron phase, we will use iron as a model material. We will ask how big a sphere of iron must be before diffusion will allow any phosphorus in it to escape, or for external oxygen atoms in silicate rock to diffuse in and turn the phosphide to phosphate, if that reaction is thermodynamically favored. We address the thermodynamics in Section 3.2 below. We assume that the iron is moving at 10 m/year, which is typical of mantle plumes [38]; other mantle convection will be substantially slower, so this is a conservative lower limit on the size of the iron bodies required.

The results of this calculation are shown in Figure 1.

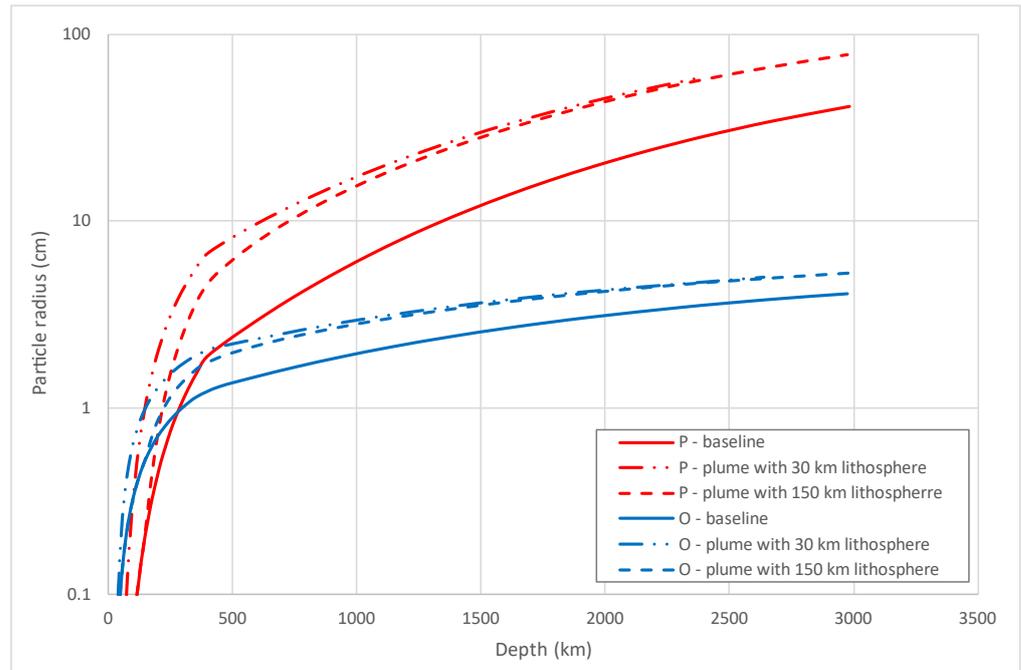

**Figure 1.** Solid pieces of iron can only be transported substantially unaltered to the surface from the deep mantle by a mantle plume if they are >1 cm radius. *X*-axis: depth (km). *Y*-axis: minimum radius for a sphere of iron to be transported from that depth to the surface, while losing no more than 50% of their phosphorus to diffusion. Plots show the temperature profile of the lithosphere, or that of a plume 300 °C hotter than the surrounding mantle [33], pooling below a 30 km [39] or a 150 km [40] thick lithosphere. Red lines show the required radius for the internal phosphorus concentration to be reduced by 50%; blue lines show the radius for the internal oxygen concentration to reach 50% of the external concentration.

Rock grains on the mantle are typically less than 1 cm [41,42], and specifically, iron grains in silicate melts are found experimentally to be typical of mm size [18]. The simple model in Figure 1 confirms that such grains will be 'frozen' in the phosphorus chemical state present at 100–200 km depth. The minerology below that depth will be erased by the traverse of the mantle. Thus, while the elemental and isotopic composition of plume volcanic melts can reflect that of the deep mantle (e.g., [43]), their chemical state will necessarily reflect the minerology of the shallow mantle or lithosphere.

There are therefore four mechanisms that could bring deep mantle material to the surface. The first, and most common, is solid state convection, during which deep mantle minerology would equilibrate to shallow minerology, erasing the deep mantle redox signature. The second is fluid flow along the surface of solid grains, which, as summarized above, would equilibrate the fluid with the redox state of the rocks over which it flowed. The third would be the rapid bulk flow of fluid through large fractures. This is the mechanism that brings rare xenoliths from the shallow mantle through the lithosphere (e.g., [32,44–49]). Fractures that propagate from the lithosphere through the deep mantle are physically unrealistic. Fractures are initiated by an overpressure of fluid in the underlying rocks. The overpressure needed to generate and sustain a fracture which is a thousand kilometers long from the deep mantle to the surface is unrealistic, and such fracturing has never been observed on Earth. Lastly, the fourth mechanism would be fluid flow from the deep mantle, which could bring meter-size fragments of phosphide mineral to the surface. This is also unrealistic as the channels between grains are a millimeter or smaller in size.

The formation of metallic phosphides by the very rapid freezing of the mantle material has happened in the fragmentation of planetisimals that formed the iron meteorites [50]. However, this event resulted in the partial or complete disruption of the parent body, which clearly has not happened to Venus.

The minerology of bulk erupted volcanic material is therefore defined by the minerology of the upper 100–200 km of the planet, which may include mantle material depending on how thick the lithosphere is. This does not preclude that some erupted material is highly reduced (e.g., [51]). On Earth, specifics of the mantle and lithosphere mineralogy result in localized ore bodies of reduced rock (e.g., [52]), and these (and rarely other rocks) can contain highly reduced xenoliths as inclusions. However, with the exception of rare xenoliths, these are the product of localized lithospheric mineralogy, not lower mantle mineralogy.

*3.2. Phosphorus Redox State in the Lithosphere and Upper Mantle*

We estimate the abundance of phosphide in volcanic ejecta using thermodynamics, extending the analysis done by [11].

The chemical state of an element in rock can be estimated from the overall redox state of elements in the rock, the temperature, and the pressure. The redox state of elements of the rock is dominated by the state of its most abundant redox active element(s), which for mantle mineralogy is usually taken to be iron. Standard redox buffers are used as references; the three cited here are quartz–iron–fayalite (QIF), which is Fe(II)/Fe(0), iron/magnetite (IM), which is $Fe_3O_4$/Fe(0), and fayalite–magnetite–quartz (FMQ), which is $Fe_3O_4$/Fe(II). The redox state of a system may be represented by its oxygen fugacity $f(O_2)$ [28], but we note that the $f(O_2)$ of a mineral assemblage changes with pressure and temperature, even if the elemental abundance in that assemblage does not change (see [28] for review).

The redox state of the Venusian lithosphere is not known, but is likely to be FMQ for two reasons. Firstly, all in situ measurements of the elemental composition of Venus' surface are consistent with it being similar to terrestrial basalt [53–55], which itself typically varies between WM and FMQ oxygen fugacity. Secondly, $SO_2$ is expected to react with surface rocks on Venus, and so, the presence of 30–300 ppm $SO_2$ below the Venusian clouds [14] requires the replenishment of the $SO_2$ by volcanism [56,57]. The <15 ppb of $H_2S$ [58] in the atmosphere suggests that the source of sulfur gases is oxidized relative to the $H_2S$/$SO_2$ couple, which requires the $O_2$ fugacity of FMQ or WM depending on assumptions about water content and temperature [11]. Under Venus surface pressure (~90 bar), $H_2S$ is slightly more abundant for any given temperature and water content than it would be on Earth (Ref [11] and the S.I. Section 2.4.2 of Ref [11]), which reinforces the idea that to achieve dominance of $SO_2$, the Venusian surface must be oxidized. Again, this does not preclude local pockets of extremely reduced (or extremely oxidized) rocks, but it does argue strongly that, as a global average, volcanic outgassing is from rocks in the WM—FMQ range of oxygen fugacity.

In principle, phosphorus could be present in iron or silicate phases of the upper mantle, and hence, in erupted magmas. We therefore calculated the partitioning of phosphorus between metal and silicate, which depends on pressure and temperature. The result in Figure 2 shows that down to 600 km depth phosphorus is overwhelmingly partitioned into the silicate phase of melts. We therefore focus on the thermodynamics of phosphorus in the silicate phase.

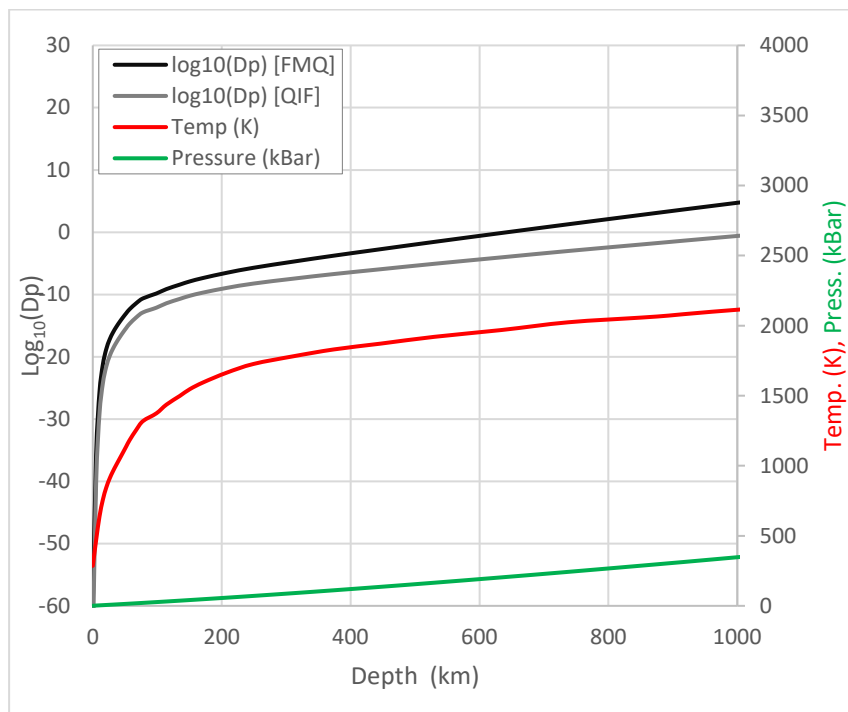

**Figure 2.** Phosphorus partitions into silicate above 1000 km depth. The partition coefficient of phosphorus $D_P$ between metallic and silicate phases predicted for the mantle shows that below 600 km (for a mantle at FMQ reduction) or 1000 km (for a QIF mantle) is the majority of phosphorus in the metallic phase as iron phosphide. $X$-axis—depth below the surface (km). Left-hand $Y$-axis: Partition coefficient $D_P$ of phosphorus between metallic and silicate phase expected at that depth. Right $Y$-axis: pressure (green) and temperature (red) at that depth.

We calculated the oxygen fugacity of the phosphate/phosphide equilibrium for all couples for iron and for all more electropositive metals than iron for phosphate and phosphide, for which thermodynamic data were available for the temperature range in the lithosphere and upper mantle. Other metals such as nickel and copper that are more electronegative than ison were not considered. If copper phosphide was present in an oxidizing environment (FMQ), it would be oxidized to copper phosphate; if present in a reduced environment (QIF), it would be displaced by more electropositive iron to form iron phosphide and copper metal. This is illustrated by the observation that meteoritic phosphide is found overwhelmingly as $Fe_3P$ (Schreibersite) and only a rare, trace mineral as $Ni_3P$ (Rhabdite). We emphasize that this does not mean that copper phosphide or other metal phosphides could not occur, just that they are unlikely to be the major phosphides in the overall shallow mantle minerology, and hence are unlikely to play a dominant role in the overall minerology of volcanism on Earth or on Venus.

The results of the thermodynamic analysis are shown in Figure 3.

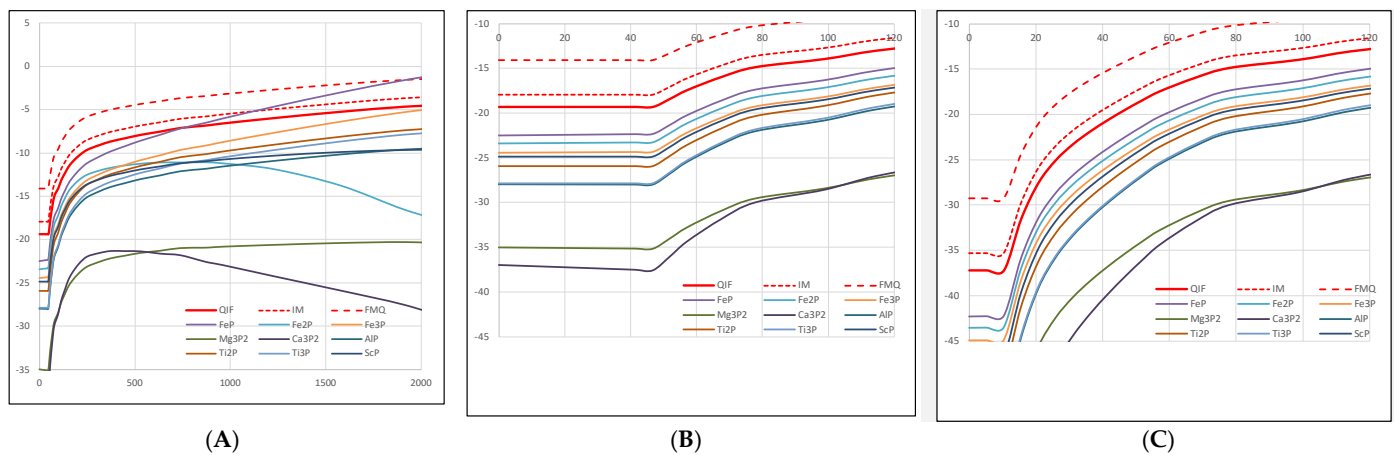

**Figure 3.** Thermodynamic analysis shows that phosphide is strongly disfavored over phosphate at lithospheric depths. *X*-axes: depth in kilometers. *Y*-axes: oxygen fugacity (log$_{10}$[f(O$_2$)]), calculated from free energy of reaction forming metal phosphide and oxygen from metal phosphate. (**A**) Full analysis to depth of 2000 km, for melt temperature of 1100 K. (**B**) Detail on the top 200 km, for melt temperature of 1100 K. (**C**) Top 200 km for melt temperature of 660 K. Iron phosphides are shown to be stable over phosphates at depths of 850 km or 1600 km if the mantle is QIF/IM or FMQ redox state respectively, but no phosphide is preferred over phosphate above 800 km in depth.

The calculation on Figure 3 clearly shows that all of the phosphates for which thermodynamic data are available are preferred over phosphides at depths shallower than 200 km. If the mantle is reduced at QIF/IM levels, then iron phosphides become preferred over phosphates at a depth of ~850 km, where the temperature is ~2000 K, which is above the melting temperature of iron phosphides [59]. This is consistent with iron phosphides being dredged from the continental 'keels' of continental cratons on Earth [60] as xenoliths. However, such xenoliths represent milligrams of material in kilograms of rock, not the bulk melt [52,61–66] (hence the name xenolith—"Foreign rock"). At lower mantle depths (Figure 3A), several phosphides are strongly favored over phosphate, consistent with such minerals being found as inclusions in mantle diamonds.

However, no phosphide is preferred over phosphate in the upper 200 km of the lithosphere. This is true whether we assume a melt temperature of 1100 K, or of 660 K (the ambient temperature at the top of the highest mountain on Venus [67]). This supports our previous calculations, which showed that it would require extraordinary circumstances for phosphorus to be present as phosphine or elemental phosphorus in near-surface rocks [11]. For an 1100 K melt and a QIF surface rock redox, and if Fe$_3$P were a dominant phosphide, then $8 \cdot 10^{-6}$ of the phosphorus in the rock would be present as phosphide. (Barringerite—Fe$_2$P—has only been found in meteorites and in pyrometamorphic rocks [68,69], and is an improbable phosphide to form in large scale volcanic melts, as iron is 40 times as abundant in the lithosphere of Earth as phosphorus, and a similar abundance ratio is expected on Venus). If the lithosphere is at FMQ redox state, and Fe$_3$P is the most likely phosphide, then phosphide will constitute only $5 \cdot 10^{-11}$ of the erupted phosphorus. We emphasize that it is the mineralogy of these near surface rocks, not the mineralogy of the deep mantle, that defines the mineralogy of volcanic erupta. We can therefore say that, even if the crust has an oxygen fugacity of QIF/IM (which, as mentioned above, is unlikely), phosphides will comprise only a very small fraction of erupted phosphorus. Therefore, to produce phosphine from plume volcanism, a substantially larger amount of rock would have to erupt than is assumed by [16]. We present a quantitative calculation of how much magma would have to erupt in Section 4 below.

In summary, phosphides might be present as <10$^{-5}$ of the phosphorus in surface (<200 km depth) melts if the melt is at the QIF redox buffer, or if they are transported from ~200 km depth as rare xenoliths. In melts at the FMQ buffer redox, which is more characteristic of terrestrial melts, phosphide would be no more than 10$^{-15}$ of the phosphorus content.

We conclude that the optimistic scenarios suggest that melts erupting on the surface could contain 10$^{-5}$–10$^{-17}$ of their phosphorus content as phosphide.

*3.3. Mechanisms of Explosive Volcanism on Venus*

Having reached the surface, phosphide minerals have to interact with acidic liquids to generate phosphine efficiently. This means that they must be fragmented into fine ash or dust and then lifted to heights of >35 km. Fragmentation is needed for two reasons; to allow the material to be lifted to high altitude and to allow it to react with sulfuric acid efficiently. Truong and Lunine's model [16] postulates that explosive volcanism lifts erupted material tens of kilometers into the Venusian atmosphere so that it can react with sulfuric acid to make $PH_3$. The most obvious site of the reaction is the sulfuric acid droplets in the clouds, but in principle phosphides, could also react with sulfuric acid vapor below the clouds, down to 35 km, below which $H_2SO_4$ is predominantly dissociated from $H_2O$ and $SO_3$ [70]. Truong and Lunine propose that ash is lifted by explosive volcanism, and use the example of the 1883 eruption of Krakatoa, which lifted several cubic miles of ash and dust into the Earth's stratosphere, as a model for such an explosive eruption. This type of volcanism contrasts with effusive eruptions, where the product is primarily lava flows which remain on the ground. We therefore next discuss whether explosive volcanism of the Krakatoa type is a realistic model for Venusian volcanism. We note that, even on Earth, explosive volcanism rarely generates ash columns that reach above 10 km. However, explosive volcanism is not likely to be common on Venus.

Explosive volcanism in this case is not just volcanism that provides a gas column that can rise to stratospheric heights, but one that also fragments a substantial fraction of the erupted magma into fine dust and ash. The observation of tropospheric $SO_2$ by Pioneer Venus [71,72] and the requirement for a source of $SO_2$ to balance reaction with crustal rocks [56,57] suggest that current volcanism is erupting $SO_2$ into Venus' atmosphere (although this itself is controversial), but is not evidence that those eruptions produce an ash column.

Explosive volcanism can only occur if erupted magma has a high content of a volatile, which drives the fragmentation of the magma, as well as the physical explosive lifting of the material into the ash cloud. On Earth, by far the most common volatile driver of explosive volcanism is water. Systems that provide explosive eruptions and ash columns are overwhelmingly seen at subduction zones or where surface water is able to interact with ascending magma (e.g., in Iceland, where volcanoes may erupt beneath the ice). This is because the subduction of the oceanic seafloor traps water, sulfates, and organics, which provide the high volatile content of resulting magmas. The 1992 Cerro Negro eruption cited by [16] is a classic example of a basaltic eruption that is nevertheless part of a subduction system [73]; in this case, the Central American Volcanic Arc resulting from the subduction zone at the western edge of the Caribbean Plate [74]. The Tarawera eruptions are also the product of subduction arc volcanism. Tarawera is part of the Taupo volcanic zone, which is subduction zone volcanism [75–77]. The 1883 eruption of Krakatoa, an unusually violent event (described as "the loudest sound heard in history" [78]) was not only the product of a subduction zone volcano but also of ocean–magma interactions which drove the final, catastrophic explosion. Similarly, the explosive eruption of the Reunion plume volcano was due to water: ocean interaction [79]. Other historically destructive explosions, such as the Minoan explosion of Thera/Santorini [80], also probably had a component of ocean–magma interactions to drive their unusually powerful eruptions and the consequent injection of ash into the high atmosphere.

Melt viscosity plays a major role in explosive volcanism, and most viscous melts on Earth, e.g., rhyolites, come from processing subducted crust [81]; the Tarawera eruptions were viscous and rhyolite-rich. However, whatever the geology, without a high content of volatiles, explosive eruption is not possible. This poses two problems for postulating explosive volcanism on Venus.

First, Venus is notoriously water-poor. Venus probably does not have Earth-like subduction zones, and certainly does not have surface oceans which could provide subducted water or interact with erupting magma. The net rate of loss of H from Venus is 1 to 2 orders of magnitude less than that from Earth [82], so the volcanic input of H into the atmosphere (as $H_2O$ or any other gas) in total must be no more than 10% of Earth's, and water-driven explosive volcanism correspondingly less frequent. Modelling suggests that explosive volcanism on Venus might occur if the water content of the melts was between 2% and 5%, depending on altitude, conduit geometry, and melt mineralogy [67,83]. Such water contents are characteristic of oceanic subduction zones on Earth [84–86]; the hotspot volcanism, which might be a better analogue for Venusian magmatism, typically has water content <0.5% [87,88]. Thus, explosive volcanism on Venus driven by water is unlikely in average geology, and hence will be rare.

The second issue with explosive volcanism as a mechanism for generating phosphide-containing ash is the consistency of having phosphide and water present at the same time in the magma. If explosive volcanism is driven by water degassing from the melt, then water must be present in the melt. Although the reaction of phosphide with water is 1000-fold slower than the reaction with acid [89] at room temperature, it is expected to be $10^{30}$ times faster at 1300 K than at 300 K, assuming $Q_{10}$ = 2. Any phosphide in the melt would therefore react with water in the melt to form $PH_3$, which is thermally unstable at melt temperatures [90]. Again, the primary driver of explosive volcanism on Earth is unlikely to deliver phosphides in large amounts.

Explosive volcanism therefore would have to be driven by a volatile other than water. Explosive volcanism could be driven by $CO_2$ (e.g., potentially the 1085 eruption of Sunset Crater, Flagstaff [91]) or $SO_2$ (e.g., the 1982 El Chichon eruption [92]—although this was a subduction volcano). This is, however, also inconsistent with the presence of phosphide. Either the bulk melt is sufficiently reduced to preserve P as phosphide, in which case C would be present as carbide or methane and S as sulfide or $H_2S$, or the melt is sufficiently oxidized for C to be present as $CO_2$ and S as $SO_2$; in which case, the P (which is more electropositive than C and S, i.e., more readily oxidized) would be present as P(+5) oxides or anions. We also note that $SO_2$ is less likely to drive explosive volcanism on Venus than on Earth, as the higher surface pressure reduces outgassing of $SO_2$ (as well as $H_2S$) (See Figure 3 in ref [83]).

(These considerations do not affect xenolith eruption, as xenoliths are chemically isolated from the melt on the melt transport timescale, but as noted above, xenoliths form an extremely small fraction of a magma.)

Chemical consistency therefore requires a melt that is very low in water but contains at least 5% of another volatile that is not $CO_2$ or $SO_2$. While such chemistry could be imagined, it would be unprecedented.

This is confirmed by the radar imaging of Venus' surface. Imaging shows very little evidence of explosive volcanism on Venus, such as pyroclastic deposits, and probable volcanic features are consistent with an overwhelmingly effusive mode of volcanism [93–95]. The same studies suggest that Venus has not undergone exceptional volcanic activity in the recent past. We conclude that explosive volcanism is both expected to be rare, and is observed to be rare, on Venus.

The alternative is that the ash clouds generated by effusive volcanism are sufficiently large to drive the required phosphide load to 35 km. Effusive volcanism does generate ash and dust, but several orders of magnitude less than explosive eruption. Studies of a number of eruptions of mantle hot-spot volcanoes show that they can produce ash clouds of several kilometers high, but that these contain 0.1–1% of the total erupted volume at most, usually much less; up to 10% of the erupted mass is produced as non-effusive material, but this reaches only tens or hundreds of meters into the air (the 'lava fountains' that characterize such eruptions) and only 2–3% or less of the ash ejecta reaches heights of 1 km or more (e.g., [96–98]). It is plausible to suggest that a much larger volcano could loft some of the plume-derived material to stratospheric heights through entrainment in the rising gas column, a possibility we address in Section 3.4 below.

We conclude that volcanism on Venus is likely to primarily be effusive, which is no more than 1–10% as efficient at delivering fine ash to high altitudes as explosive eruptions, and probably would deliver only trivial amounts.

*3.4. Ash Plume Generation on Venus*

Several factors mitigate against any volcanic eruption on Venus being as efficient at delivering ash to >35 km altitudes as it would be on Earth, regardless of its mineralogy. These are discussed in detail in [95]. Specifically, ash generation is less efficient, and plumes are likely to be smaller.

To generate fine ash, magma must be fragmented from bulk material into very small fragments. Regardless of magma chemistry, high surface pressure will suppress ash generation. At higher surface barometric pressure, gases will exsolve less and into smaller volumes, which will suppress rapid bubble expansion, leading to lower degrees of fragmentation.

Eruption itself only projects magma a few hundred meters into the atmosphere. To rise further, ash must be entrained in a rising column of hot gas. The lift driving volcanic plume ascent into planetary atmospheres comes from heat exchange from the fragmented magma to the atmosphere, heating the gas, and thereby producing buoyancy. The efficiency of this heat exchange is a function of magma fragmentation, which, as noted above, is likely to be lower on Venus than on Earth. The resulting buoyant driving force will depend in part on the temperature lapse rate of the atmosphere, which is less on Venus than on Earth [72]. Head and Wilson [95] estimates that plumes on Venus will be ~1/3 the height those generated by similar eruptions on Earth. Extremely high volatile fractions (>12% $SO_2$ for example) are needed to create a 50 km height ash cloud under Venus conditions [95].

Thus, although we adopt a simple scaling that effusive volcanism produces 10–100 times less ash than explosive volcanism, the above considerations point to this being highly conservative in the case of Venus, where the real decrease in fine ash production may be more like > 1000 times less than for a terrestrial explosive eruption.

*3.5. Other Criteria for a Volcanic Source*

We note that, in the literature, data on the hydrolysis of phosphides cited by Truong and Lunine relate to aqueous solutions of sulfuric acid, not concentrated acid. Concentrated sulfuric acid is an oxidizing agent, not a hydrolyzing reagent; noted in [11]. We also believe that Truong and Lunine overestimate the mantle phosphorus content. We present isotope record data in [11] that suggests that, outside very limited regions, the mantle phosphorus abundance is likely to be ~0.2 wt.%. Further analysis of >2000 rock samples from Earthchem [99] suggests an average phosphorus abundance of 0.074 wt.% in terrestrial igneous rock samples.

*3.6. Required Rate of Phosphide Volcanic Eruption*

We will briefly address the amount of mineral that is required to produce the flux of phosphine necessary to maintain 1 ppb gas at cloud level in the atmosphere. The required $PH_3$ production rate has been dealt with in detail in [11], so we will only summarize the argument here.

Ref. [11] provides a detailed photochemical analysis of the rate of $PH_3$ production necessary to sustain a 1 ppb $PH_3$ level in the clouds. Ref. [16] simplifies this, and estimates the phosphine production rate required to explain the putative phosphine signal by

$$R_P = \frac{M_{PH_3}}{t_{PH_3}}, \qquad (7)$$

where $M_{PH_3}$ is the mass of phosphine in the 53–61 km detection layer reported by [1] and $t_{PH_3}$ is the photochemical lifetime of $PH_3$ at 53–61 km. Ref. [16] further estimates $t_{PH_3}$ by the transport timescale for a molecule to cross ~10 km, which is in the order of magnitude of the distance a molecule, would need to traverse to reach the bottom or top of the Venusian clouds, where its lifetime is short due to photochemical destruction [1].

The ref. [16] calculation underestimates the required production flux of phosphine, for three reasons:

1. It assumes that phosphine is only present from 53–61 km. However, the very same diffusive processes invoked to transport phosphine to the upper atmospheres mean that phosphine must be present at altitudes beyond 53–61 km as well. In other words, this approach underestimates $M_{PH_3}$.

2. It overestimates the lifetime of high-altitude phosphine. The eddy diffusion coefficient increases with altitude in the Venusian atmosphere, meaning that the use of $10^3$ cm$^2$ s$^{-1}$ underestimates the phosphine destruction rate. Therefore, it overestimates $t_{PH_3}$ for PH$_3$ at z > 63 km, which must exist due to the diffusive transport.
3. It assumes that phosphine is only destroyed after transport. In fact, phosphine is continually destroyed throughout the Venusian atmosphere due to radical sinks [1].

When we attempt to account for these factors by running a full 1D photochemical model, we find a required PH$_3$ production rate of 26 kg/s, compared to 1 kg/s for [16] (to produce 1 ppb PH$_3$). We note that this probably under-estimates the necessary PH$_3$ production rate, as PH$_3$ is oxidized efficiently by liquid concentrated sulfuric acid [13], a sink not accounted for in any models for lack of kinetic data.

This is not to discount the usefulness of such calculation; indeed, we executed a variant of it ourselves ([11] and Section 3.2.4.1 in ref. [11]). This calculation is valuable because it involves few parameters and is therefore less vulnerable to the considerable photochemical uncertainties. Further, such calculation has the philosophical advantage that it seeks to explain solely the detected phosphine, and thus, in a sense, provides the most conservative approach (easiest for known mechanisms to explain). However, it has the disadvantage that, in order to be true, it requires transport in the Venusian atmosphere to behave in ways atmospheric transport is not known to behave. Especially, it requires PH$_3$ not to exist outside of the ~53–61 km layer, while the same diffusion which transports the PH$_3$ to be destroyed should also result in PH$_3$ existing outside the 53–61 km layer. For the [16] explanation to function, Venus must host a unique atmospheric vertical transport pattern which sequesters PH$_3$ in the cloud deck.

For the purposes of this paper, we adopt a range of fluxes; a lower limit of the 1 kg/s adopted by Truong and Lunine [16] as a conservative value, and a higher limit of the [11] value of 26 kg/s.

## 4. Discussion

Truong and Lunine [16] revisited the idea of mantle plume volcanism as a source of the PH$_3$ tentatively detected on Venus. We have built on this hypothesis quantitatively in terms of atmospheric chemistry, mantle mineralogy, and eruption physics, and chemistry. As a base case, [16] estimate that 1 kg/sec of phosphides needs to be delivered to clouds, and this requires 0.03–0.15 km$^3$/year of material containing 1% phosphorus, 5% of this is phosphides, and a $t_{PH_3}$ = 10$^9$ s. In Table 2, we summarize the factors needed to correct for a more complete model of phosphorus geochemistry, assuming that the lithosphere is reducing (QIF oxygen fugacity). We conclude that the actual amount of magma that needs to erupt is ≥21,600 km$^3$/year. Even the Deccan and Siberian trap emplacements (which might have been up to 10 km$^3$/year at their peak [16,100,101]), fall short of the required rate of phosphide injection by at least three orders of magnitude. This therefore represents an unprecedented level of volcanic activity.

**Table 2.** Summary of factors required for Truong and Lunine [16] base model.

| Factor | Comment | Section | Multiplier |
|---|---|---|---|
| Phosphine lifetime | $t_{PH_3}$ could be as specified by Truong and Lunine or full photochemical model | 3.6 | 1–26 |
| Abundance of erupted phosphide, assuming QIF f($O_2$) | <$10^{-5}$ of phosphorus likely to be phosphide within 100 km of surface | 3.1, 3.2 | >6000 |
| Delivery of phosphide to clouds | Effusive volcanism at least 10-fold less efficient at producing fine ash as explosive volcanism | 3.3, 3.4 | $10^1$–$10^2$ |
| Abundance of phosphorus | Most likely 0.08%, not 1% | 3.5, [11] | 12 |
| Overall scale-up of [16] estimate required to meet $PH_3$ production requirement | | | ≥7.2·$10^5$ |
| Volume of volcanic eruption required, based on Truong and Lunine base case of 0.03–0.15 km³/year | | | ≥26,100 km³/year |

21,600 cubic kilometers is quite a lot of magma; enough to build a volcano the size of Olympus Mons every 138 years [102], or to resurface the planet to a depth of 1 km in 22,000 years. It clearly cannot be from one volcanic plume; if this were a single volcanic conduit through which magma was flowing at 10 m/sec (an exceptionally high velocity for terrestrial volcanoes [103,104]), then such a flux would require a 'dike' with an area of ~68,500 square meters, or 12 American Football fields, an unprecedented event. The result summarized in Table 2 assumes an extremely conservative model of surface geochemistry, volcanic mechanics, and of phosphine photochemistry. If we instead adopt the more physically realistic photochemical model of [1,11], then the rate of phosphide delivery to the clouds is 26 times that assumed here, and the volcanic eruption rate is similarly higher at ~560,000 km³/year. If we assume that the lithosphere has an oxygen fugacity nearer FMQ than QIF, then the required eruption rate is several orders of magnitude higher still. If we adopt a more realistic rate of ash cloud generation, then the required eruption rate is again increased.

We conclude that large-scale volcanism is a plausible source for $PH_3$ on Venus if

a) The atmospheric structure and gas transport on Venus are such that $PH_3$ (and only $PH_3$) is retained solely in the lower cloud layer, and does not diffuse upwards or downwards, and

b) The Venusian lithosphere is substantially more reduced than is expected from lander data and atmospheric chemistry, and

c) Volcanism is currently erupting tens of thousands of cubic kilometers of magma onto the surface per year, and

d) Venusian volcanism is unexpectedly efficient at generating high altitude clouds of fine magma ash.

None of these requirements is unphysical, but all are unexpected and can be tested by observation. We note that the unexpected atmospheric physics can only apply to $PH_3$, and not to other minor gases (which are adequately modelled using conventional mixing processes).

Very high volumes of volcanic eruption should have altered the surface over hundreds of square kilometers since radar observations of Venus began in the 1970s, so systematic searches for such changes could test for the reality of such a very large-scale volcanism. Additionally, the thermal imprint of such large-scale volcanism would be expected to be visible in the cloud layer as a hot spot whose source co-rotated with the planet and not with the clouds.

There remains the possibility that very high volumes of magma happened to be erupting in 1978 (when Pioneer Venus sampled the atmosphere), and in 2017–2019, when the Greaves et al. [1] observations were made, but eruptions have been quiescent for most of the intervening 40 years. This would require only one or two volcanoes the size of the Hawaiian island complex [105] to be accumulated since Venus' surface observations began under our most optimistic assumptions, but would still point to Venus entering a new phase of volcanism unlike that of the recent past.

## 5. Conclusions

The presence of $PH_3$ in Venus' atmosphere is unexpected and invites explanation. Volcanism is only an adequate explanation if our understanding of basic diffusion processes in Venus' atmosphere, the minerology of Venus' surface, and of the mechanisms of ash cloud formation are incorrect, and if Venus is currently undergoing a major eruption event of a magnitude greater than any seen on Earth in the Phanerozoic, and different from the volcanism regime in Venus' recent past. It has been suggested that Venus has undergone such major resurfacing as global resurfacing ~300 Mya, in possibly a single catastrophic event that lasted less than 10 My [17]. If Venus were starting such an event now, it could explain the presence of phosphine in the clouds. Such an exciting possibility should be explored.

**Supplementary Materials:** The following are available online at www.mdpi.com/xxx/s1: Spreadsheet containing the following sheets: Buffers (table of standard buffer fugacity as a function of depth), Fugacity Calculation (table of free energy of formation of minerals and calculation of oxygen fugacity for selected phosphide/phosphate buffers), References (literature references for the thermodynamic data), Read Me (details of how the calculations are performed in this spreadsheet).

**Author Contributions**: Conceptualization, W.B. and O.S.; methodology, W.B., O.S., S.R.; writing—original draft preparation, W.B.; writing—review and editing, all authors; funding acquisition, S.S., P.B.R., O.S. All authors have read and agreed to the published version of the manuscript.

**Funding:** We thank the Change Happens Foundation for funding. S.R. acknowledges the funding from the Simons Foundation (495062). P.B.R. thanks the Simons Foundation for funding (SCOL awards 599634).

**Institutional Review Board Statement:** Not applicable.

**Informed Consent Statement:** Not applicable.

**Data Availability Statement:** All relevant data is provided in Supplementary Materials to this article.

**Acknowledgments:** We are very grateful for the constructive comments provided in review.

**Conflicts of Interest:** The authors declare no conflict of interest.